# Continuous-wave, singly resonant OPO at 3 μm

**S. T. Lin[1], Y. Y. Lin[1], Y. C. Huang[1,*], A. C. Chiang[2], and J. T. Shy[3]**

[1]*Institute of Photonics Technologies, Department of Electrical engineering, National Tsing-hua University, Hsinchu 30013, Taiwan*

[2]*Nuclear Science and Technology Development Center, National Tsing-hua University, Hsinchu 30013, Taiwan*

[3]*Department of Physics, National Tsing-hua University, Hsinchu 30013, Taiwan*

*\*Corresponding author: ychuang@ee.nthu.edu.tw*

**Abstract:** We report a continuous-wave, singly resonant optical parametric oscillator (SRO) at 3.3 μm by using a MgO:PPLN crystal as the gain medium and a Yb-fiber laser at 1.064 μm as the pump source. At 25-W pump power, the SRO generated 7.4 and 0.2-W powers at 1.57 and 3.3 μm, respectively. The 3.3-μm SRO has a single-longitudinal-mode output with a ~5-MHz linewidth. We observed thermally induced optical bistability in the SRO due to the slight absorption of the 3.3-μm wave in the MgO:PPLN crystal. Pump depletion was clamped at 60% while spectral hole burning was observed in the depleted pump.

©2008 Optical Society of America

**OCIS codes:** (190.4970) Parametric oscillators and amplifier; (190.1450) Bistability.

## 1. Introduction

Wavelength-tunable, narrow-line, continuous-wave (CW) laser sources in the 3-4 μm region are useful for applications such as environmental sensing and monitoring. CW, singly resonant optical parametric oscillators (SROs) using periodically poled lithium niobate (PPLN) crystals as the parametric gain medium have been demonstrated with high conversion efficiency and narrow linewidth over a broad tuning range in the 3-μm wavelengths by resonating a ~1.5 μm signal in the resonator [1,2]. A ring SRO is capable of generating a single-longitudinal-mode (SLM) output at the resonant wavelength while inheriting the spectral characteristics of the pump source in the nonresonant wave [3]. Therefore, for a pump laser at 1 μm and a resonant wave near 1.5-1.7 μm, a delicate single-frequency pump laser is usually needed to ensure a single-frequency output near 3-4 μm. Although directly resonating the idler wave at 3-4 μm in a PPLN-based SRO could generate a SLM output in the mid-infrared by using an economical multi-longitudinal-mode (MLM) pump, the potential challenge associated with the high pump threshold resulting from material absorption and transverse-mode mismatch renders it relatively unexplored. Here we report, for the first time to the best of our knowledge, a single-frequency CW SRO resonating at 3.3 μm pumped by an economical broadband Yb-fiber laser at 1.064 μm. As will be shown below, interesting phenomena such as optical bistability and thermal guiding are observed in this SRO.

In this work, we used a 5 mol.%-doped MgO:PPLN crystal with a domain period of 30.7 μm (HC Photonics) as the parametric gain medium of the 3.3-μm SRO. The crystal is 50-mm long ($x$), 3-mm wide ($y$), and 1-mm thick ($z$). Congruent lithium niobate doped with MgO was known to resist photofractive damage, and exhibit better power and temperature stability in an optical parametric oscillator [4]. However, lithium niobate is also known to have increased absorption at wavelengths near 3 μm, which is expected to increase the oscillation threshold of an SRO. The high intracavity power of the resonant wave at 3 μm would also cause absorption heating to the lithium niobate crystal and potentially lead to thermal focusing, dephasing, aberration, and even material fracture [5,6]. Moreover, the spatial overlapping between the idler and pump waves is worse than that between the signal and pump waves due to the vast difference in the idler and pump wavelengths, which could further increase the oscillation threshold of a SRO resonating the 3-μm idler wave. In our experiment, although the 3.3 μm resonant wave is not located in the strong absorption region of lithium niobate, we indeed observed self heating and thermally induced beam distortion. As will be shown below, when the circulating idler power is sufficiently large, the thermally induced optical guiding in the lithium niobate crystal surprisingly improves the parametric efficiency.

**2. Experimental Setup**

The schematic of our bow-tie ring SRO is showed in Fig. 1, where the four reflecting mirrors are made of infrared-grade fused silica. The pump laser is a linearly polarized Yb-fiber laser (IPG YLM-25-LP) at 1064 nm, producing 25-W CW power in a 1-nm or 265-GHz linewidth. The pump beam was polarized along the crystallographic z direction of the MgO:PPLN crystal and mode-matched to the SRO cavity by using a 150-mm focal-length lens. The waist radius of the pump beam at the center of the MgO:PPLN crystal is 90 μm. The 30.7-μm period of the MgO:PPLN crystal permits the first-order quasi-phase matching of the pump, signal and idler waves at 1.064, 1.57 and 3.3 μm, respectively, at a crystal temperature of 50 °C. The two end apertures of the PPLN crystal were optically polished and coated with 0.5, 0.2, and 14% reflectance at the pump, idler, and signal wavelengths, respectively.

To tune the OPO output wavelengths over a wide range, the MgO:PPLN crystal was installed in a controlled oven with a temperature resolution of 0.1 °C. The total cavity length of the ring SRO is 500 mm and the cavity-mode radius at the center of the MgO:PPLN crystal is 100 μm. The two curved mirrors, $M_1$ and $M_2$, are separated by 140 mm with a radius of curvature of 100 mm on the reflecting surfaces. The other two mirrors are separated by 100 mm with flat reflecting surfaces. The four cavity mirrors all have >99.8% reflectance over a wavelength range of 3200-3400 nm, and >97% transmittance over a wavelength range of 1550-1650 nm and at the pump wavelength.

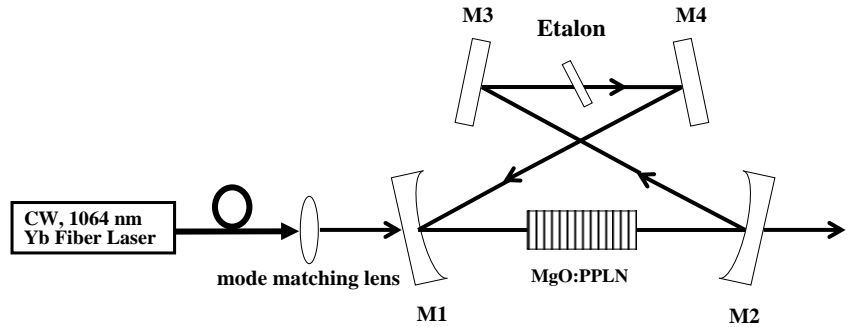

Fig. 1. The configuration of the 3.3-μm SRO pumped by an Yb-fiber laser at 1064 nm. The bow-tie ring cavity consists of two curved mirrors with 100-μm radius of curvature and two flat mirrors highly reflecting the resonant wave at 3.3 μm. The mode matching lens focuses the pump beam to the center of a MgO:PPLN crystal. The etalon helps lock the cavity mode and tune the output wavelength of the idler wave.

## 3. Result and Discussion

Figure 2 shows the measured signal power versus the pump power. At 25-W pump power, 7.4-W signal power at 1.57 μm was emitted through M2 and 50-mW idler power at 3.3 μm was emitted through each of the four cavity mirrors. Owing to the diffraction and absorption loss of the 3.3-μm resonant wave in the MgO:PPLN crystal, the ~9-W oscillation threshold of this SRO is about twice the threshold of a similar SRO resonating at the signal wavelength [2]. The filled and open dots denote the output signal powers when the pump power was increased from 0 to 25 W and back to 0 W, respectively.

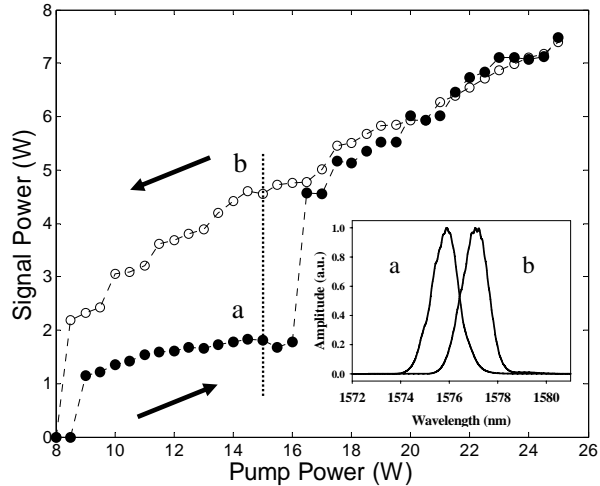

Fig. 2. Measured output signal power versus the pump power of the SRO. The filled and open dots denote the measured signal power when the pump power was varied from 0 to 25 W and back to 0 W, respectively. The inset shows a shifted signal spectrum for the two bistability states at 15-W pump power.

It is interesting to see that a sudden increase of the signal power from 1.7 to 4.2 W, accompanied by a sudden change in the output beam profile, occurs at a pump power of ~16.5 W. Decreasing the pump power does not trace back the SRO output power measured during the increasing phase of the pump power. There are two stable states for this SRO when the pump power is between 8.5 and 16.5 W. Compared with the low-power state (State a), the high-power state (State b) has a higher conversion efficiency and a lower threshold value (around 8.5 W). The inset shows the signal wave spectra of the two states at 15-W pump power and 65 °C oven temperature. The linewidths of these two signal waves are both 1.3 nm or 160 GHz. The central wavelength of the high-power state is 2 nm longer than that of the low-power state, which can be explained by a 4 °C temperature increase in the MgO:PPLN crystal for the high-power state. Therefore the hysteresis loop in Fig. 2 is associated with thermally induced optical bistability. From the raised crystal temperature and sudden change of the spatial-mode profile, we speculate a threshold pump power of 16.5 W to induce the thermal guiding in the MgO:PPLN crystal. With the optical guiding, the increased conversion efficiency and lowered oscillation threshold are attributable to the improved spatial overlapping between the pump and idler waves in the MgO:PPLN crystal.

Figure 3 shows the pump depletion versus the pump power and the number of times above threshold. Since the oscillation thresholds for the two bistability states are different, to be conservative, we use the higher of the two values or 9 W for calculating the number of times above threshold. After the pump power reaches the oscillation threshold, the pump depletion of the low-power state jumps to about 25% but decreases to 20% at 1.9 times above threshold. However, the pump depletion is increased to and clamped at 50~60% after the pump power exceeds the thermal-guiding threshold or 2 times above the oscillation threshold. Because of the thermal guiding, 50~60% pump depletion is maintained over almost the whole range of the high-power state (State b). The clamped pump depletion of our SRO is somewhat different from the monotonically increased ones previously reported for narrow-line (~2.2 GHz and ~5 kHz) pumped ring SROs resonating the 1.5-μm signal wave [1,2]. The inset in Fig. 3 shows the measured power spectrum of the depleted pump laser at 20-W input power. The spectral hole burning in the depleted pump indicates that the broad linewidth of our Yb-fiber laser, 1 nm or 265 GHz, is the main cause of the clamped pump depletion. From the depleted pump spectrum, the optimal pump-laser linewidth for the SRO should be within ~60% of the 1-nm linewidth of our Yb-fiber laser, which is consistent with the measured 160-GHz signal linewidth in the inset of Fig. 2.

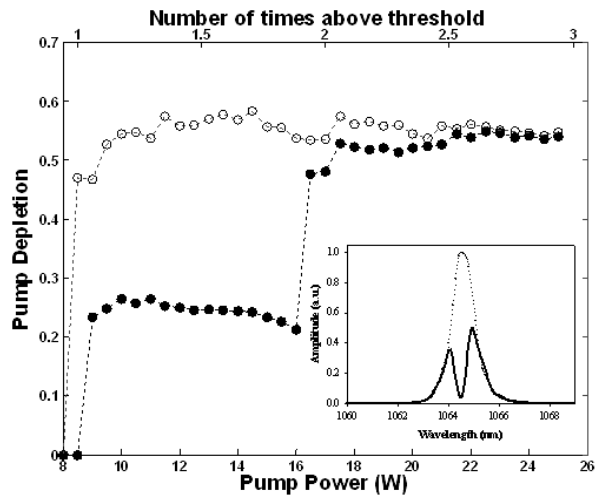

Fig. 3. Pump depletion versus the pump power and the number of times above threshold. States a (filled dots) and b (open dots) denote the two bistability states. The pump depletion is clamped at 50-60% when the pump power exceeds the thermal guiding threshold at 16.5 W. The inset shows spectral hole burning in the depleted pump, which explains the clamped pump depletion.

To determine the power stability of the 3-μm SRO, we measured the output signal power with a thermal detector when the pump power was fixed at 20 W. To lock the cavity mode and reduce the spectral noise, a z-cut, 780-μm-thick, double-side polished, uncoated lithium niobate plate was used as an intra-cavity etalon (free spectral range, FSR = 90 GHz). Figure 4 shows 0.6% and 1.1% rms signal power fluctuations over 10-minute and 2-hour (inset) spans of time. The overall signal power was decreased by ~5% from that in Fig. 2 due to the insertion of the etalon. At the full 25-W pump power, we measured a single-mode linewidth of ~5 MHz at the idler wavelength by using a scanning confocal Fabry-Perot spectrometer (FSR= 0.75 GHz, Finesse > 600).

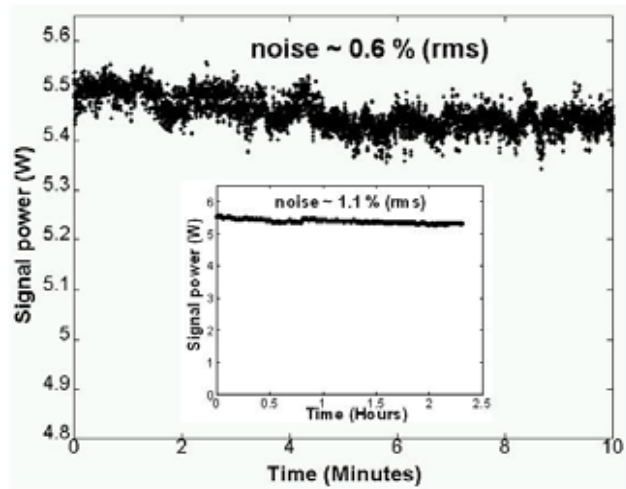

Fig. 4. The noise characteristics of the 3.3-μm SRO. With the intra-cavity etalon, the short-term (over 10 minutes) and long-term (over 2 hours) rms noise are 0.6 and 1.1% (inset), respectively.

## 4. Summary

In summary, we have successfully demonstrated a CW, SLM, SRO at 3.3 μm by using a MgO:PPLN crystal as the parametric gain medium and an economical 1-nm linewidth Yb-fiber laser as the pump source. This optical parametric oscillator is, to the best of our knowledge, the first CW SRO with a resonant wavelength longer than 2 μm. We also reported the first observation of thermally induced optical bistability in a mid-infrared SRO. The slight absorption of the 3-μm resonant wave in the MgO:PPLN crystal, although increasing the pump threshold, helps mode-overlapping and parametric conversion in the MgO:PPLN crystal when the pump power reached a thermal-guiding threshold of 16.5 W. The measured spectral hole burning in the depleted pump suggests further increase of the parametric efficiency with a pump linewidth less than 160 GHz.

## 5. Acknowledgements

This work was supported in part by NTHU under Code 96N2534E1 and in part by NDIDF under Code 96A0105N6. Y. C. Huang's email address is ychuang@ee.nthu.edu.tw.